# FRAMEWORK FOR INTEGRATING ZERO TRUST IN CLOUD-BASED ENDPOINT SECURITY FOR CRITICAL INFRASTRUCTURE

**Shyam Kumar Gajula**
USA.

**ABSTRACT**

*Cyber threats have become highly sophisticated, prompting a heightened concern for endpoint security, especially in critical infrastructure, to new heights. A security model, such as **Zero Trust Architecture** (ZTA), is required to overcome this challenge. ZTA treats every access request as new and assumes no implicit trust.*

*Critical infrastructure like power plants, healthcare systems, financial systems, water supply, and military assets are especially prone to becoming targets for hackers and phishing attacks. This proposes a comprehensive framework for integrating tailored ZTA into organizations involved in managing sensitive operations.*

*The paper highlights how the ZTA framework can enhance compliance, enabling continuous protection, thereby reducing attack surfaces. This paper aims to address the gap that exists in applying ZTA to endpoint management within cloud environments for critical infrastructure.*

**Statement of proposed framework:**

This paper proposes a multi-layered framework for integrating ZTA into critical infrastructure. The layers include identity-centric controls, continuous device assessment, and micro-segmentation to create a strong defense model that protects critical systems against sophisticated cyberthreats in cloud-based environments.

# 1. Introduction

## 1.1. Background of traditional security models:

In the traditional security model, the company network is the castle, while the firewalls and Virtual Private Networks (VPN) act as the moat. It's a perimeter-based approach, called the 'castle and moat,' with the focus on defending the perimeter to keep out unauthorized users. These models worked well when employees and devices or systems were on-premises. As work has shifted from secure premises to open networks at homes and public places, the risk of security compromise has also increased.

Perimeter-based defense is proving to be inadequate for fending off cyber criminals. Recent reports show that [ransomware incidents increased 17.8% year over year,](#) and encrypted attacks rose 10.3% year over year. Organizations are increasingly being exposed to malware attacks due to unpatched vulnerabilities, forcing them to move toward ZTA as quickly as possible.

## 1.2. Rise of cloud computing and remote work- a new threat landscape:

Cloud computing can be traced back to ARPANET, the early internet of the 1960s, which laid the framework for eventual breakthroughs by allowing access to information and applications in remote devices. By the early 2000s, cloud computing had become a household name in the workplace as businesses outsourced work, and the world started getting more connected.

More recently, with COVID-19, businesses accelerated worldwide into a new paradigm. Work from home or remote work became the new norm for organizations to keep functioning. Cloud computing has progressed significantly since then, enabling businesses to access powerful computing capabilities without making expensive investments. However, with the growing dependence on cloud-based services, companies are becoming susceptible to cyberattacks and malware threats.

There's no guarantee that data stored in the clouds is safe. Also, firms that provide these services cannot be fully trusted to adequately secure customers' data. This requires a robust safety framework to prevent cyberattacks on sensitive data, especially critical infrastructure, which is the backbone of society.

**1.3. Why critical infrastructure requires stricter security controls:**

Critical infrastructure, as mentioned before, is the backbone of an effectively functioning society and a nation as a whole. It includes physical and virtual assets, networks, and systems in sectors such as military, energy, water supply, sewage disposal, transportation, food and agriculture, emergency services, banking, telecommunication, utility, etc.

Critical infrastructure may include industrial control systems (ICS), including supervisory control and data acquisition (SCADA) systems that are used to automate industrial processes. Cyberattacks against one of these systems can [compromise the integrity of other vital systems,](#) like a cascading effect, as often these infrastructure systems are connected and interdependent. When multiple sectors are affected, it can cripple large portions of society dependent on these systems, bringing essential services to a standstill.

The financial implications of such attacks are also serious. The average cost of downtime per minute can be anywhere from $5,000 to $10,000 in large oil and gas and manufacturing organizations, according to Gartner. It's not difficult for nation-states to extort abysmal amounts of ransom from rival countries if they have an upper hand in cyberwarfare.

Governments and corporations must be one step ahead of such threats to ensure their critical infrastructure is safe from them. This is where Zero Trust Architecture (ZTA) can play a major role in ensuring critical infrastructure is protected from malicious entities seeking to undermine the sovereignty of a state or nation.

Zero Trust Architecture (ZTA) is based on zero-trust principles and has no trusted networks or locations. ZTA prohibits lateral movement within an enterprise, preventing easy access to applications once a user enters the network. It achieves this by continuous monitoring of user movement, contextual authentication, such as reviewing users' location, device identities, and authentication tokens.

**1.4. Objectives and scope of the paper:**

The objective of this paper is to advocate and implement a 'never trust, always verify' mentality, especially when it comes to critical infrastructure. It also addresses challenges faced by agencies seeking to implement ZTA. It introduces a framework to apply ZTA effectively to safeguard vital systems against persistent security threats from malicious actors within and outside an organization.

We understand that attacks on critical infrastructure are becoming more real each day as global tensions escalate and nations rise against nations on every continent of the world. It is time to stay 10 steps ahead with stricter, robust security controls owing to the nature of the threat and the sophistication of our adversaries.

## 2. Literature Review
### 2.1. Evolution of Endpoint Security
#### 2.1.1. Traditional endpoint protection vs. EDR/XDR

Endpoint security has transitioned from basic to sophisticated protection solutions. Antivirus software was heavily relied upon and was designed to detect known threats, thereby failing against new ones.

On the other hand, Endpoint Detection and Response (EDR) and Extended Detection and Response (XDR) have become game changers as they offer real-time threat detection and incident response.

Cloud-based endpoint management tools like Microsoft Intune, JAMF, and CrowdStrike support remote device oversight in distributed environments, further transforming the landscape.

### 2.2. Zero Trust Architecture (ZTA)
#### 2.2.1. Origin and evolution

ZTA was introduced by John Kindervag in 2011 and has become increasingly critical to implement in critical infrastructure, considering the speed with which cloud-based applications, multi-cloud environments, and Internet of Things (IoT) have taken center stage. Users now demand quick and direct access to resources from anywhere in the world to collaborate and remain productive. ZTA actually makes this possible without compromising on security.

With [NIST's SP 800-207](#) publication, ZTA has gained recognition and structure. Its core tenets include:

- **Never trust, always verify** – Continuous monitoring and validation of resource usage is critical to detect unusual behavior. User authenticity must be verified by using multifactor authentication, device health check, and application whitelisting.
- **Enforce Least Privilege Access** – This principle restricts the users' rights to data, services, and applications through which they perform their authorized functions. Enforcement is done using just-in-time (JIT), just-enough access (JEA), and granular access controls. This helps minimize exposure or damage related to compromised user accounts or insider threats.

- **Micro-segmentation/Assume security breach** – ZTA assumes that security breaches are inevitable and can originate inside or outside the organization's perimeter. ZTA aims to minimize the extent of the breach when it occurs. This requires micro-segmenting sensitive resources, continuous monitoring of user and device behavior, using end-to-end encryption, and implementing robust mechanisms for incident response and recovery.

ZTA's relevance to critical infrastructure across various sectors lies in its ability to address threats from insiders and from compromised credentials. Strict identity checks and micro-segmentation minimize the effect a potential breach can cause within a sector. Any industry can effectively implement ZTA across a variety of use cases like finance, healthcare, retail, energy, government, and defense.

## 2.3. Security in Critical Infrastructure

Power grid – A simple power cut for maintenance purposes cripples our daily routine. Considering that, the scope of disruption a cyberattack on the power grid can cause cannot be underestimated. Bad actors can breach SCADA systems and feed instructions to transformers and generators to perform actions that may lead them to overheat or overload and explode, causing fires. Such an attack was [experienced by Ukraine](#) on its power grid in 2015, which not only led to widespread power disruptions but also ensured that restoration efforts would be delayed by corrupting the software.

Banks and financial institutions – Millions of people trust their life savings, perform transactions, and even trade with banks and financial institutions. Everything in this sector is trust-based, and that trust can be breached in a moment by malicious actors. Billions can be erased from the market by hackers in just minutes by manipulating trading algorithms, sending stock prices crashing. An attack on core banking systems like SWIFT could freeze global business money transfers, bringing the economy to a standstill.

Healthcare – Ransomware can encrypt a hospital's Electronic Health Record (EHR), preventing emergency access for patient care. Attackers can also manipulate connected medical devices (IoMT) like pacemakers and infusion pumps, turning them into death machines. The [US FDA recalled](#) half a million Abbott Laboratories pacemakers in 2017 due to fears that hackers could manipulate the device and run batteries down or alter the patient's heartbeat.

Defense – Several government agencies and private contractors research, develop, and manufacture military weapons and systems. Hackers can easily steal classified documents and blueprints for sensitive technology like missiles, aircraft, anti-aircraft technology, etc. They can compromise the security of personnel working in sensitive departments or stationed at sensitive locations, putting their lives at risk.

Critical manufacturing – Many industries manufacture critical equipment parts and machinery for critical sectors. These transactions include sensitive drawings, materials, software, and technology, which can be stolen by hackers. In addition, an attack on such facilities can cripple timely production or even manipulate software to create faulty parts that can severely affect the functioning and readiness of critical sectors.

## 2.4. Minimum Regulatory Landscape

### 2.4.1. North American Electric Reliability Corporation – Critical Infrastructure Protection (NERC CIP)

These are a highly prescriptive, [legally enforceable set of requirements](#) for the North American bulk power system. Fines can reach millions of dollars per day per violation. The standards mandate specific controls like defining a clear Electronic Security Perimeter (ESP), performing background checks on authorized personnel, enforcing strict access control to prevent unauthorized logins, etc.

### 2.4.2. Health Insurance Portability and Accountability Act (HIPAA)

HIPAA requires healthcare organizations to [secure patient data](#) by implementing administrative safeguards like policies and procedures, physical safeguards like secure data storage, and technical safeguards like encryption and audit logs.

### 2.4.3. NIST Cybersecurity Framework

It has [five core functions](#) – Identify, Protect, Detect, Respond, and Recover. For example, for a water utility infrastructure, this means: **Identifying** their critical assets like chemical dosing systems and pressure monitors; **Protecting** them with firewalls and access controls; **Detecting** anomalies like an unauthorized attempt to log into a control system; **Responding** by isolating the system if breached; **Recovering** by restoring operations from secure backups.

Common cyber threats in CI environments

Some common threats faced by cloud security include:
- **Insider threats** – A threat that comes from within the organization could be from a former or current employee, or any other person with direct access to the company network. In Oldsmar, Florida, an upset engineer used their credentials to change chlorine levels in a water treatment plant. Another operator, noticing the change, corrected it on time.

- **Cyberattacks** – These threats, such as phishing, malware, SQL Injections, DoS, DDoS, come from cyber criminals or hackers who attempt to access a computer network for malicious purposes such as stealing, altering, or destroying data. In 2021, hackers attacked the Colonial Pipeline by encrypting its IT billing and administration network. The pipeline had to be shut down, leading to fuel shortages and a declaration of national emergency.

- **Zero-day exploits** – These threats target vulnerabilities in unpatched software and operating systems. Even with a high-end cloud configuration, an attacker can exploit zero-day vulnerabilities to access the environment.

- **Advanced Persistent Threats (APT)** – These are sophisticated and sustained attacks in which the attacker establishes an undetected presence within the network to steal sensitive information over a period of time. For example, the TRISIS/TRITON malware discovered in a Saudi Arabian petrochemical plant was designed to specifically target the plant's safety systems, which were meant to shut down the plant in the event of an emergency.

## 3. Research Gaps and Problem Statement

### 3.1. Lack of tailored frameworks addressing endpoint-centric ZTA for CI

There is a lack of a specific framework tailored for ZTA application to endpoint management within cloud environments for critical infrastructure. Endpoints have become the new perimeter due to remote accessibility, and generic ZTA models are unable to provide sufficient security for unique security demands like those for the energy and defense sectors. This paper aims to propose a framework, especially in this context.

#### 3.1.1. Challenges in integrating ZTA principles into legacy infrastructure

A significant challenge is integrating modern ZTA principles with legacy Operational technology (OT) used in critical infrastructure like ICS and SCADA. These systems are difficult to replace and are not able to deal with the current threat landscape. It is difficult to apply the ZTA "never trust, always verify", without disrupting essential national services.

**3.1.2. Gap in interoperability and implementation standards**

The NIST SP 800-207 provides a structure for ZTA; however, there is no clear interoperable blueprint for consistent application of ZTA across different sectors, hindering effective and uniform adoption.

**4. Proposed Conceptual Framework**

**4.1. Components of the Framework**

Identity-centric access control – This component makes user identity the primary control point by enforcing the principle of least privilege using multi-factor authentication and just-in-time access. This ensures access to only what's required for the user to perform their work.

Device posture assessment – This component continuously assesses endpoints to validate their security health before granting access by checking for malware, updated patches, compliance, etc., preventing suspicious devices from connecting to critical infrastructure.

Micro-segmentation – This component creates granular security zones around each application and workload, preventing lateral movement of attackers, limiting the effective radius of the breach.

Continuous authentication and authorization – 'Once in,' doesn't mean 'always in' for this component. Access is continuously evaluated in real-time based on user behavior, context, and device health.

Policy engines and enforcement points – The policy engine makes access decisions in a dynamic manner, while agents and gateways apply these policies to the resource edge.

Monitoring and telemetry – Logs and data are collected from all access requests and endpoints, providing crucial information to detect suspicious activity and deploy a rapid incident response.

Cloud-native control plane for endpoint management – All security policies across distributed endpoints are orchestrated by a centralized cloud-based platform, enabling consistent management for remote work environments.

**4.2. Architectural Layers**

The proposed framework has four distinct architectural layers, with each layer performing a specific function and working together to create a dynamic and intelligent security boundary around each access request.

### 4.2.1. Presentation Layer (User/Device Interaction)

This layer is the entry point into the ZTA environment, representing the various endpoints that have become [the new security perimeter](). Here is where the user gains access to a critical infrastructure using their device from any location, such as an engineer trying to access the maintenance logs of a power plant using their tablet or a financial analyst trying to access a trading platform from their home. This layer's function is to present the security context of the user and the device they are using for verification by subsequent layers. This is the initial access request that triggers the "never trust, always verify" workflow.

### 4.2.2. Access Control Layer (Policy Enforcement)

This layer acts as the gateway between users and resources, and is the core decision-making and enforcement layer of the framework. This is where the policy engine and enforcement points are housed. A request from the presentation layer is first intercepted here, and the intelligence layer is put to work to gather context such as the user's identity, device compliance, and hints of suspicious behavior. Based on these inputs, it makes real-time decisions based on pre-defined security policies. Here, the Identity and Access Management (IAM)systems enforce authentication and authorization, and continuous authentication is applied. For OT or SCADA systems in critical infrastructure, a policy enforcement point would act as a secure proxy, isolating the critical system while enforcing modern, identity-based access controls.

### 4.2.3. Intelligence Layer (Threat Detection, Logging)

This layer provides the critical data and analytics needed for the access control layer to make informed decisions. Using EDR/XDR tools, it gathers a vast amount of data from endpoints, network traffic, and application logs. It is responsible for monitoring and telemetry. The security information and event management (SIEM) system correlates events to detect anomalies, insider threats, and the [subtle tactics of Advanced Persistent Threats]() (APTs). The device posture assessment occurs here. This layer alerts the access control layer if an unusual pattern, such as unauthorized access outside working hours, is detected, and then dynamically adjusts access rights or terminates the session.

### 4.2.4. Infrastructure Layer (Cloud, Hybrid Systems)

This is where all the protected resources are and encompasses the entire hybrid and multi-cloud environment of a critical infrastructure organization. It includes everything from cloud-native applications and databases to the legacy ICS and SCADA systems that control physical processes in critical sectors. Micro-segmentation is applied here, creating isolated environments and preventing lateral movement. It is an important recommendation for securing industrial control systems.

## 4.3. Framework Diagram

**High-level visual of your Zero Trust-based endpoint security architecture.**

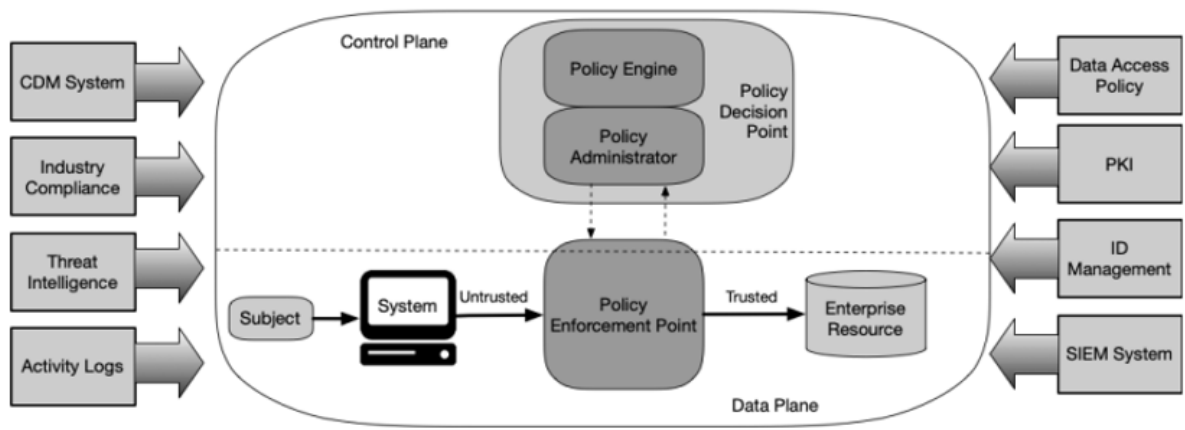

Figure 2: Core Zero Trust Logical Components

Link: https://nvlpubs.nist.gov/nistpubs/SpecialPublications/NIST.SP.800-207.pdf

## 5. Conclusion

This paper addresses the serious security gaps in critical infrastructure and proposes a comprehensive, endpoint-centric, zero-trust framework. It suggests shifting the paradigm from implicit trust to continuous verification across its different security layers, which offers a resilient and adaptable defense against sophisticated cyber threats facing critical infrastructure running the sensitive assets of nations.

We aim to focus our research on developing pilot implementations in simulated critical infrastructure environments to test the framework's efficacy and performance. Work is also required to create measurable milestones for progress, and standardized ZTA implementation step-by-step guides to the unique regulatory and operational needs of specific sectors like energy or healthcare.

For the successful implementation of ZTA, it is important to implement it in phases. Organizations should begin with foundational elements like strong identity-centric controls using IAM and MFA, along with EDR/XDR for device posture assessment. This creates immediate value before implementing the full network micro-segmentation, which is a far more complex task.

Retrofitting ZTA principles into legacy OT and SCADA systems without disrupting critical operations is a significant challenge. In addition, the ZTA framework's effectiveness is also dependent on overcoming the traditional mindset to adapt to the "never trust" mindset.

The implementation is a technically complex exercise requiring integration of different security tools within a cohesive intelligence layer.